# Study of the spring and autumn daemon-flux maxima at the Baksan Neutrino Observatory

E. M. Drobyshevski[1] and M. E. Drobyshevski[1,2]

[1] *Ioffe Physico-Technical Institute, Russian Academy of Sciences, St.Petersburg, 194021Russia;*
 *E-mail: emdrob@mail.ioffe.ru*
[2] *Astronomical Department, Faculty of Mathematics and Mechanics, St.Petersburg State University, Peterhof, 198504  St.Petersburg, Russia; E-mail: miked@mail.ru*

**ABSTRACT**

Detection of daemons in low-background conditions in September 2005 and March 2006 has provided supportive evidence for the expected to occur at that times maxima in the flux of daemons with $V \approx$ 10-15 km s$^{-1}$, which hit the Earth from near-Earth, almost circular heliocentric orbits (NEACHOs). The ability of some FEU-167-1 PM tubes with a thicker inner Al coating to detect directly (without a scintillator) daemon passage through them has also been demonstrated, an effect increasing 100-fold the detector efficiency. As a result, the daemon flux recorded at the maxima was increased from ~$10^{-9}$ to ~$10^{-7}$ cm$^{-2}$s$^{-1}$. At the maxima, two phases in the observed flux can be discriminated. The first of them is associated with objects which catch up with the Earth in moving in outer NEACHOs and cross it. The intensity and direction of the flux during this phase which lasts about two weeks depend on the time of day and latitude of observations (therefore, synchronous measurements in the Northern and Southern Earth's hemispheres are desirable). In the second phase, where the flux consists primarily of few objects captured into geocentric, Earth-surface-crossing orbits (GESCOs) during the first phase, the daytime and latitude dependence becomes less pronounced. The experiments suggest an explanation for the fairly poor reproducibility of our earlier ground-level measurements (subtle differences in PMT design, varying radon background etc.). All the experimental results thus obtained either support the conclusions following from the daemon paradigm or find a simple interpretation within it.

# 1 INTRODUCTION

The properties of Dark Electric Matter Objects, daemons, which made possible their detection by the scintillation technique are in line with the expected characteristics of elementary black holes with $m \approx 3\times 10^{-5}$ g, Planckian particles, if they carry a negative electric charge $Ze \approx -10e$ corresponding to their mass (e.g., Markov 1966). Among them is capture of nuclei in matter, with their excitation and ejection of electrons and nucleons (Drobyshevski 1997), and subsequent daemon-stimulated disintegration in the remainder of the nucleus of one proton after another with an interval of ~1 μs, which makes possible capture of another nucleus (Drobyshevski 2000). (At a velocity $V \approx 10$ km s$^{-1}$, disintegration of a nucleus comes to an end within a path of ~10 cm, and the capture of another nucleus in air, in ~1-3 mm.) Of considerable significance is also a high penetrability of the daemon permitting it to cross the Sun and the Earth with a very weak deceleration, a property that enables daemons populating the Galactic disk to become captured and accumulate in heliocentric orbits including NEACHOs and, subsequently, in GESCOs, which contract in a relatively short time to disappear finally inside the Earth.

Daemon transfer to NEACHOs is most probable in the zones where the projection of the Earth's orbital velocity onto the solar apex direction is maximal. This is why the daemon flow undergoes a semiannual periodicity. These zones lie close (by pure chance) to the equinoxes (Drobyshevski 2005).

Detection of daemons at the Ioffe Physico-Technical Institute, RAS, St.Petersburg, was carried out by Drobyshevski (2002) and Drobyshevski *et al.* (2003a) with a thin (~10 μm) ZnS(Ag) layer deposited on the underside of two horizontal polystyrene plates, 0.5×0.5 m$^2$ in area and 4 (or 1) mm thick, spaced by 7 cm and arranged at the center of a tinned-iron cube, 51 cm on a side. Eight such modules were employed altogether. Each plate is viewed by its FEU-167 PM tube, whose output is fed into a dual-trace digital storage oscillograph. The time shift $\Delta t$ between the beginnings of the scintillator signals provides a judgment of the object velocity. There are two types of signals. Signals of the first type, with a long (~2 μs) flat maximum 2-2.5 μs distant from the beginning, are scintillations caused by heavy nonrelativistic particles, e.g., α-particles (Heavy Particle Scintillations, HPSs); the other type, initiated, in particular, by cosmic rays (these are signals with $\Delta t = 0$, occurring sometimes simultaneously in several modules) and characteristic of intrinsic PMT noise, have a steep (~1.5 μs) leading edge ending in a sharp maximum (Noise-Like Signals, NLSs). We focused out attention on double events with HPSs in the top channel, whose output triggered the oscillograph sweep.

Because most of the double events originate from background produced by cosmic rays and natural radioactivity, the events of interest to us are isolated by a statistical analysis of the distribution of the number of events $N(\Delta t)$ in the time shift $\Delta t$ separating them. Significant results were obtained with the first four modules almost immediately, in March 2000, when a maximum in $N(\Delta t)$ was observed within the $+20 < \Delta t < 40$ μs bin. The statistical significance of this $+30$-μs peak was 2.85$\sigma$. Its small width corresponds to the small spread in the velocity of the objects striking the Earth from NEACHOs. One more March observation, we thought, and the significance would rise to 4$\sigma$. Because, however, of the continuous changes of system parameters in hope to reveal novel effects the detector efficiency decreased. And only summation of data amassed in the months of March 2000, 2003, 2004, and 2005 raised the significance of this peak to 99.99% (Drobyshevski 2006; thereafter Paper I). Although in these years we have detected a seasonal variation with a period $P = 0.5$ yr (Drobyshevski *et al.* 2003b) and some indications of a September maximum similar to the March one, it became clear that some important parameters are still eluding us.

The ground-level observations performed in March 2005 suggested (for more details, see Paper I) that some FEU-167 PM tubes are capable of generating themselves, without scintillator, a signal when crossed by a daemon. These seemed to be PMTs with a larger (up to ~1 μm) thickness of the inner Al coating of their photocathode section. This thickness is usually not specified because it does not affect the photometric characteristics of a PM tube, but, according to the original specification, should be ≈0.1 μm.

Indeed, in passing a distance of 4-6 cm in the vacuum cathode section of a PM tube, about 50% of daemons should reduce the charge $Z_n$ of the nuclei captured outside in air to the level where $|Z_n| - |Z| < 1$, or, colliding in vacuum with no new nuclei, even to $|Z_n| - |Z| \approx -(4 \div 6)$. As the daemon enters now the internal, electrically conducting PM coating and captures there with a high probability a nucleus with the concomitant ejection of nucleons and hundreds of (refilling) electrons, it generates in the PM tube a measurable signal. This behavior was observed in the bottom PM tubes of three out of eight modules (these are FEU-167-1 PMTs with envelopes made of potassium-containing glass). Attributing the origin of the 30-μs peak in these modules to the above effect, the effective detector area decreases markedly, so that the calculated daemon flux rises to $f_\oplus \approx 0.5 \times 10^{-7}$ cm$^{-2}$s$^{-1}$ (Paper I).

Whence it immediately follows that a detector should operate even without the bottom scintillator, if the PM tube itself acts as a sensitive element. It is this consideration, as well as a desire to carry out the September and March maxima in well controlled, low-background conditions, that have motivated the present experiments in the Baksan Neutrino Observatory (BNO). The specific features of the detector used are described in Section 2. Sections 3 and 4 list and discuss the data obtained in September, while the March data will be discussed in Section 5 and 6. The measurements to be analyzed provide compelling evidence not only for the existence of both the latitude and, possibly, diurnal variations within the September and March maxima in the flux of low-velocity daemons from the outer NEACHOs but for the possibility of using PM tubes as primary daemon detectors too, as well as stress the need of monitoring the content of radon in air in ground-level observations.

## 2 DESCRIPTION OF THE DETECTOR AND ITS SETUP

We brought from the Ioffe PTI to the BNO only detector module #3 (see Paper I). The bottom PM tube in the module was screened by two layers of 6-μm-thick Lavsan film coated on one side with 0.05-μm-thick Al layer. We left only one polystyrene plate, 4 mm thick, with its ~6-mg cm$^{-2}$-thick ZnS(Ag) scintillator layer *face up*, and it was viewed by the top PMT from a distance of 22 cm. Heretofore, the plates in our modules were always turned with the ZnS(Ag) layer down (the only exclusion having been made for the upper 1-mm-thick plates in modules #21-#24 in the measurements of March 2005 (Paper I)). Now the plate had to be turned in order to put two transparent 50-μm-thick Lavsan films on the ZnS(Ag) layer. In addition, the scintillator corners were screened with black paper, so that the scintillator area viewed by the PMT amounted to 2100 cm$^2$. The films protected the scintillator from direct incidence of α-particles of radon and of radiative products of its decay contained in air. In addition, the module was continuously blown from below by nitrogen evaporated from a Dewar flask (12 cm$^3$s$^{-1}$ flow rate).

These measures permitted us, starting from September 3, 2005, to reduce the average HPS triggering rate from the top PMT from the initial level of ~0.75 s$^{-1}$ down to one trigger in ~60 s, at our standard oscillograph trigger level of ≈2.6 mV (events with $U_1 \geq 2.8$ mV were processed). NLS-initiated trigger signals come 10-12 times less frequently than the HPS-produced ones. In the period from September 3 to 23, during 436 h of live time operation, we recorded 27600 trigger signals, with 290 of them having been double events (one event per about 95 triggers), out of which, in their turn, 25 events were triggered by NLSs on the top trace, including 7 events with $\Delta t \leq 0.4$ μs (which adds up to 0.35 events per day). This did not come as a surprise, because the module was installed at a depth of 400

MWE, and NLSs with Δt ≈ 0 are produced, as we now know, primarily by cosmic ray muons crossing the PMT dynode blocks (Paper I).

In the ground-level experiments performed in St.Petersburg (its latitude is 60°) in March 2005 (Paper I), during the total time of 413 h ≈ $1.5 \times 10^6$ s, trigger signals from module #3 ($3 \times 10^5$ signals altogether) followed with an average interval of ~5 s. One event was recorded per 28 trigger signals. The total number of double events recorded was about $10^4$. They included 69% NLSs with Δt ≈ 0, 21.4% NLSs with Δt ≠ 0, and 9.6% with HPSs on the first trace. Thus, the number of NLS events with Δt ≈ 0 caused by cosmic rays at the PTI is ~$10^3$ times that obtained at BNO. Also, the rate of detection of double events with HPS recorded on the upper trace with the same module #3 at BNO (13 per day) is only one fourth that at PTI (55 events per day), which suggests a substantial effect of the radon background on ground-level measurements. All this makes the advantages of carrying out experiments at BNO only too obvious.

## 3  RESULTS OF THE SEPTEMBER EXPERIMENT. SOME SURPRISES

The final adjustments completed, the module was put in operation and exposed from September 3 to 23, 2005. The complete list of main results, including the $N(\Delta t)$ distributions accumulated by the time of arrival of a possible event of daemon fall from the NEACHO is presented in Tables 1 and 2.

**Table 1**. $N(\Delta t)$ distributions (-100 < Δt < 100 μs) accumulated from September 3 ($11^h$) to September 18 ($17^h$) 2005 (here and subsequently, the times are reckoned from local astronomical midnight). Ten bins, each 20 μs wide, are centered at the Δt specified in the Table. The $N(\Delta t)$ distributions are given for the time of recording of an event in the -30-μs bin. $\Sigma$ is the total number of events in $N(\Delta t)$; $N_{-30}$ is the number of events in the -30-μs bin; $\{\sigma\} = (N_{-30} - 0.1\Sigma) \cdot N_{-30}^{-1/2}$ refers to statistical significance of the -30-μs maximum; $\{N\}_9 = (\Sigma - N_{-30})/9$ is the average number of 'background' events per one bin (excepting the −30 μs bin).

| Event № | Sept. days | -90 | -70 | -50 | -30 | -10 | 10 | 30 | 50 | 70 | 90 | $\Sigma$ | $\{\sigma\}$ | $\{N\}_9$ |
|---|---|---|---|---|---|---|---|---|---|---|---|---|---|---|
| 249471 | 3,825 | 1 | 0 | 1 | 1 | 0 | 1 | 0 | 1 | 0 | 0 | 5 | 0,50 | 0,44 |
| 250237 | 4,487 | 2 | 1 | 1 | 2 | 0 | 2 | 1 | 1 | 1 | 2 | 13 | 0,49 | 1,22 |
| 251259 | 5,417 | 4 | 1 | 1 | 3 | 1 | 4 | 1 | 1 | 1 | 3 | 20 | 0,58 | 1,89 |
| 251656 | 5,707 | 4 | 1 | 1 | 4 | 1 | 4 | 2 | 1 | 1 | 3 | 22 | 0,90 | 2,00 |
| 252241 | 6,044 | 6 | 2 | 1 | 5 | 1 | 4 | 2 | 2 | 2 | 3 | 28 | 0,98 | 2,56 |
| 253473 | 6,897 | 6 | 2 | 1 | 6 | 1 | 5 | 3 | 2 | 3 | 6 | 35 | 1,02 | 3,22 |
| 253524 | 6,931 | 6 | 2 | 1 | 7 | 1 | 5 | 3 | 2 | 3 | 6 | 36 | 1,28 | 3,22 |
| 253549 | 6,956 | 6 | 2 | 1 | 8 | 1 | 5 | 3 | 2 | 3 | 6 | 37 | 1,52 | 3,22 |
| 254583 | 7,801 | 7 | 3 | 1 | 9 | 2 | 6 | 3 | 4 | 4 | 7 | 46 | 1,47 | 4,11 |
| 255192 | 8,246 | 7 | 3 | 2 | 10 | 2 | 8 | 3 | 4 | 4 | 8 | 51 | 1,52 | 4,56 |
| 256517 | 9,237 | 9 | 3 | 3 | 11 | 3 | 8 | 4 | 4 | 4 | 9 | 58 | 1,57 | 5,22 |
| 256729 | 9,397 | 9 | 3 | 3 | 12 | 3 | 9 | 5 | 4 | 4 | 10 | 62 | 1,67 | 5,56 |
| 257069 | 9,612 | 10 | 4 | 3 | 13 | 4 | 9 | 5 | 4 | 5 | 10 | 67 | 1,75 | 6,00 |
| 257695 | 10,073 | 11 | 4 | 5 | 14 | 4 | 9 | 5 | 4 | 6 | 10 | 72 | 1,82 | 6,44 |
| 257951 | 10,243 | 11 | 4 | 5 | 15 | 4 | 9 | 5 | 4 | 6 | 10 | 73 | 1,99 | 6,44 |
| 259242 | 11,102 | 12 | 5 | 5 | 16 | 5 | 10 | 5 | 6 | 8 | 11 | 83 | 1,92 | 7,44 |
| 259362 | 11,182 | 12 | 5 | 5 | 17 | 5 | 10 | 5 | 6 | 8 | 11 | 84 | 2,09 | 7,44 |
| 259530 | 11,312 | 12 | 5 | 5 | 18 | 5 | 10 | 5 | 6 | 8 | 12 | 86 | 2,22 | 7,56 |
| 260149 | 11,725 | 13 | 6 | 7 | 19 | 6 | 10 | 8 | 6 | 9 | 12 | 96 | 2,16 | 8,56 |
| 261715 | 12,851 | 17 | 7 | 9 | 20 | 7 | 15 | 8 | 8 | 11 | 12 | 114 | 1,92 | 10,44 |
| 262937 | 13,733 | 18 | 7 | 9 | 21 | 10 | 18 | 8 | 10 | 13 | 12 | 126 | 1,83 | 11,67 |
| 265114 | 15,335 | 20 | 9 | 10 | 22 | 12 | 18 | 12 | 12 | 15 | 15 | 145 | 1,60 | 13,67 |
| 266215 | 16,052 | 21 | 10 | 11 | 23 | 14 | 18 | 12 | 15 | 15 | 15 | 154 | 1,58 | 14,56 |
| 266349 | 16,151 | 22 | 10 | 11 | 24 | 14 | 18 | 13 | 15 | 15 | 16 | 158 | 1,67 | 14,89 |
| 266396 | 16,191 | 23 | 10 | 11 | 25 | 14 | 18 | 13 | 15 | 15 | 16 | 160 | 1,80 | 15,00 |
| 270012 | 18,647 | 28 | 13 | 17 | 25 | 15 | 23 | 16 | 18 | 24 | 19 | 198 | 1,04 | 19,22 |

**Table 2.** $N(\Delta t)$ distributions accumulated from September 18 ($18^h15'$) to September 23 ($12^h45'$) 2005 after the interchange of the PMTs in the detector module (the top PMT had a screened photocathode, and the bottom PMT viewed the 2500-cm² ZnS(Ag) layer and triggered the oscillograph upper trace). The $N(\Delta t)$ distributions correspond to recording a new event in the +30-μs bin; $\{\sigma\}$ is calculated for the bin centered at $\Delta t = +10$ μs (i.e., for the $0 \leq \Delta t \leq +20$ μs bin); the values of $\{N\}_9$ were derived with subtraction of the number of events in the +30-μs bin and normalized against the 2100-cm² area of the ZnS(Ag) layer.

| Event № | Sept. days | $N(\Delta t)$ | | | | | | | | | | $\Sigma$ | $\{\sigma\}$ | $\{N\}_9$ |
|---|---|---|---|---|---|---|---|---|---|---|---|---|---|---|
| | | -90 | -70 | -50 | -30 | -10 | 10 | 30 | 50 | 70 | 90 | | | |
| 271683 | 19,727 | 2 | 1 | 1 | 1 | 1 | 2 | 1 | 2 | 2 | 1 | 14 | 1,33 | 1,21 |
| 273471 | 20,929 | 4 | 2 | 2 | 2 | 1 | 8 | 2 | 5 | 2 | 2 | 30 | 1,77 | 2,61 |
| 274775 | 21,953 | 7 | 2 | 3 | 2 | 2 | 8 | 3 | 5 | 4 | 4 | 40 | 1,41 | 3,45 |
| 274862 | 22,018 | 7 | 2 | 3 | 2 | 2 | 8 | 4 | 5 | 4 | 4 | 41 | 1,38 | 3,45 |
| 276666 | 23,478 | 7 | 3 | 4 | 5 | 5 | 11 | 5 | 8 | 5 | 5 | 58 | 1,57 | 4,95 |

The most essential finding is that the bottom FEU-167-1 with screened photocathode does indeed respond to its traversal by daemons, as could be expected from the ground-level experiments at the PTI. Indeed, the conclusion that the $N(\Delta t)$ distributions, treated by the $\chi^2$ criterion, cannot be represented by a constant has a significance of ~90-99%.

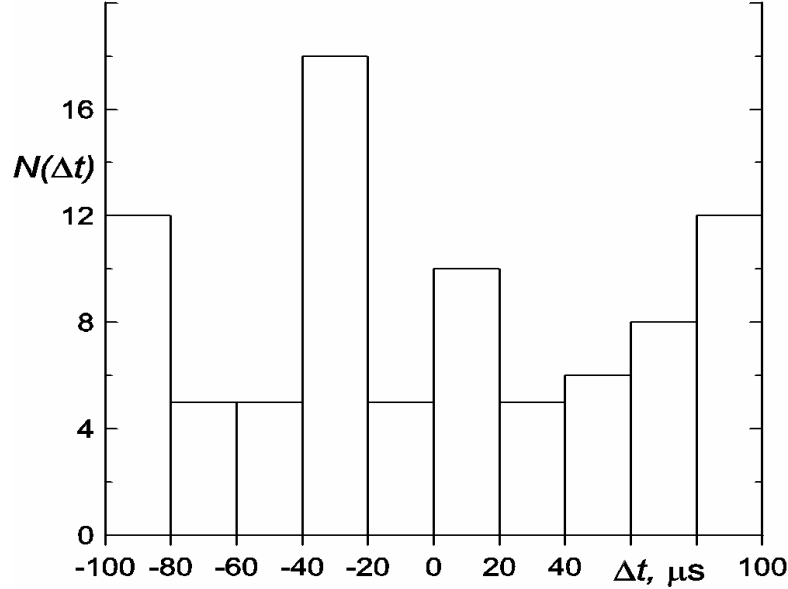

**Figure 1.** $N(\Delta t)$ distribution of double events, with HPSs on the top PMT viewing the ZnS(Ag) scintillator, in the time shift $\Delta t$ between the beginnings of the signals. The photocathode of the bottom PM tube is screened with Al foil. The observations were run from September 3 ($11^h$) to September 11 ($7^h30'$) 2005; 86 events altogether. Statistical significance of the maximum in the -30-μs bin $\{\sigma\} = (N_{-30} - 0.1\Sigma) \cdot N_{-30}^{-1/2} = 2.2\sigma$.

One more striking observation. Figure 1 displays, in our opinion, the most significant $N(\Delta t)$ distribution. The expected peak is seen to be confined not in the $+20 > \Delta t > 40$ μs bin, but rather in an exactly opposite bin $-20 > \Delta t > -40$ μs! Said otherwise, the events recorded were generated by objects that had crossed the Earth while practically retaining their velocity characteristic of falling from the NEACHOs onto the opposite side of the Earth. The statistical significance of the peak is as high as $2.2\sigma$, which is substantially more than we expected and should be assigned, as we believe, to the reduced (radon) background. If such data had been obtained at BNO in a standard four-module setup like this was done at St.Petersburg, the significance would have certainly been about $4\sigma$.

## 4 DISCUSSION OF THE SEPTEMBER RESULTS

Figure 2 presents graphically the variation of some measured parameters with time, which was drawn using Tables 1 and 2.

We see plotted here the variation both of the number of events $N_{-30}(t)$ in the -30-μs bin, and of the average number of "background" events $\{N(t)\}_9$ per one of the remaining nine 20-μs bins. The latter is approximated with a confidence level 99.5% by a straight line $1.27 \cdot t - 5.77$, which corresponds, on the average, to 12.7 events per day in $-100 < \Delta t < 100$ μs range.

We see plotted here the variation both of the number of events $N_{-30}(t)$ in the -30-μs bin, and of the average number of "background" events $\{N(t)\}_9$ per one of the remaining nine 20-μs bins. The latter is approximated with a confidence level 99.5% by a straight line $1.27 \cdot t - 5.77$, which corresponds, on the average, to 12.7 events per day in $-100 < \Delta t < 100$ μs range.

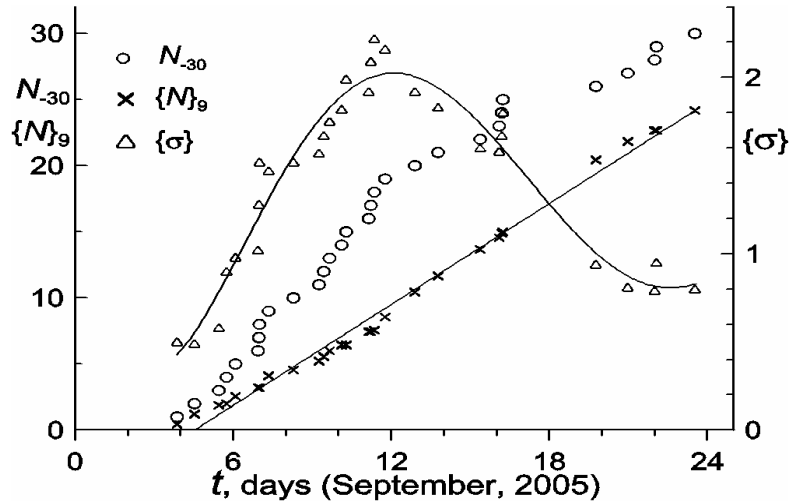

**Figure 2.** Behavior of some parameters with time $t$ of the observations run in September 2005. $N_{-30}(t)$ is the number of events in the -30-μs bin of the $N(\Delta t)$ distribution; $\{N(t)\}_9$ is the number of background events occurring, on the average, in each of the 9 bins, excluding the -30-μs bin ($\{N(t)\}_9 = 1.27 \cdot t - 5.77$ approximation has a C.L. = 99.5%); $\{\sigma\}$ is statistical significance of the -30-μs bin maximum in the $N(\Delta t)$ distribution.

We readily see that the rate of rise of the number of events in the -30-μs bin is initially quite high, noticeably higher than that of background events, but ten days after the beginning of observations the rise stops (Fig. 3). It appears that we have indeed, as expected, arrived exactly at the September maximum of the NEACHO daemon flow, when the Earth passes through the crowding zone of these orbits, where they cross one another and the Earth's orbit. It would hardly be possible to maintain that we have observed the very beginning of this flow, and, therefore, we did not approximate the data in Fig. 3 with a probability integral. Nevertheless, if we assume the dependence of flux intensity on time to be a Gaussian, its halfwidth should be not less than 3 days, and the maximum flux becomes $f_\oplus \approx 1.0 \times 10^{-7}$ cm$^{-2}$s$^{-1}$, which practically coincides with the March flux $f_\oplus \approx 0.5 \times 10^{-7}$ cm$^{-2}$s$^{-1}$ derived from ground-level observations in 2005 but is at least an order of magnitude lower than the real flux, if we allow for the low efficiency of the scintillator part of the detector (Paper I). The duration of the flow, only ~10 days, which corresponds to ~10° displacement of the Earth along its orbit, is unexpectedly short. Further experimental and theoretical studies would be needed to find an explanation for this.

Figure 2 plots also $\{\sigma(t)\}$, the behavior with time of the statistical significance of the -30-µs peak in $N(\Delta t)$ distributions obtained on different days in September. On September 11 (for $N_{-30}$ = 18), the statistical significance reaches $\{\sigma\}$ = 2.2$\sigma$ (C.L. = 97%). We calculate the statistical significance of a peak in the simplest way possible, namely, divide the excess of the number of events in the peak over the arithmetic mean of the ten bins by the square root of the number of events in the peak been. This is enough to see how improbable is a purely stochastic appearance of such an overshoot. More sophisticated approaches to calculation of the statistical significance (Taylor 1982), for instance, by division of the peak excess over the weighted mean by the square root of the sum of the squared weighted mean error and of the number of events confined in the peak, while yielding a somewhat higher statistical significance (2.5$\sigma$ in place of 2.2$\sigma$ in the above example), carry a slight flavor of scholastics. Application of these methods could hardly be justified here, because the numbers of events in neighboring bins of the $N(\Delta t)$ distribution are not statistically independent, which follows from the daemon hypothesis. The $N(\Delta t)$ distribution varies continuously with time, as we believe, e.g. because of some daemons transferring from NEACHOs to GESCOs, i.e., through their diffusion in the velocity-time (and, see below, Earth's latitude) space. This brings about a broadening of the -30-µs peak and appearance in the $N(\Delta t)$ distribution of statistically significant side lobes corresponding to GESCOs with velocities of up to 3-5 km s$^{-1}$ (see Fig. 1, as well as Drobyshevski *et al*. 2003a). This is why after September 17-18 one observes even a decrease of the $N_{-30}(t)$ - $\{N(t)\}_9$ difference (see Fig. 3 and Tables 1 and 2).

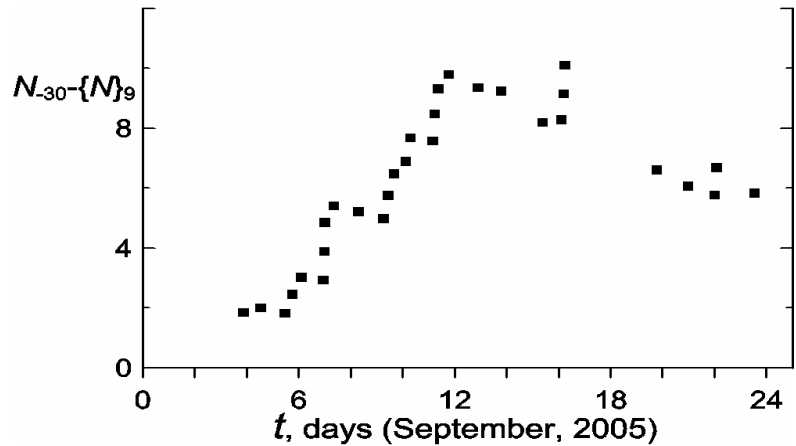

**Figure 3.** Excess in the number of events $N_{-30}$ in the -30-µs bin above $\{N\}_9$ background plotted vs. time. $N_{-30}$ - $\{N\}_9$ for $t$ after September 18.75 was calculated as $N_{+30}$ - $\{N\}_9$ and by summing the data in Tables 1 and 2. After September 17, no daemons with velocities 10(11.2)-15 km s$^{-1}$ have been observed manifestly.

Consider now the maximum in the 0 < $\Delta t$ < 20 µs bin. In the first and last days of the experiment it even competes with the -30-µs peak. Treated from the standpoint of the daemon hypothesis, it corresponds to a velocity >15 km s$^{-1}$, but its intensity and position (downward flux with no excess events in the neighboring +30-µs bin) are not clear. We are inclined to assign its appearance to excitation of scintillations in the ZnS(Ag) layer by some short-lived (~10$^{-5}$ s) radioactive or isomer nuclei, which decay with emission of a particle (for instance, a neutron) entering the oppositely arranged screened PM tube with a corresponding delay.

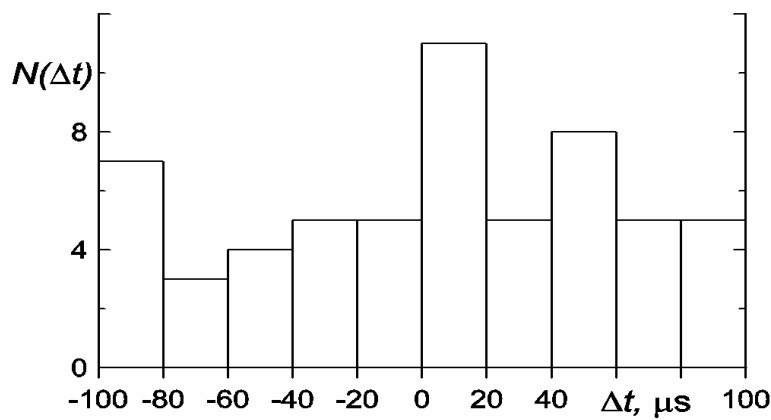

**Figure 4.** $N(\Delta t)$ distribution after 4.7 day exposure of the detector with PMTs interchanged on September 18.75 (see text and Table 2). 58 events altogether. The maximum in the +10-μs bin has $\{\sigma\} = 1.57\sigma$. PMT interchange did not change the position of this invariably present maximum (see Fig. 1 and Table 1, and Sec.5 below) adjoining $\Delta t = 0$ from the right, which evidences its non-daemon origin.

To check this assumption, we waited until the -30-μs peak practically stopped to grow, and, starting from September 18, $17^h 50'$ (and to the end of the experiment, September 23, $12^h 45'$) interchanged the PM tubes: the PMT viewing the scintillator (the scintillator area seen from below is 2500 cm$^2$) and triggering the oscillograph was placed below, and the PM tube with the photo-cathode screened by Al foil was placed at the module top. The $N(\Delta t)$ distribution obtained in such an arrangement was expected to become exactly opposite to the previously obtained $N(\Delta t)$ relative $\Delta t = 0$; indeed, particles propagating downward should now have yielded signals with $\Delta t < 0$, and vice versa. This PMT rearrangement did not change the position of the maximum under study (Table 2 and Fig. 4), it remained in the $0 < \Delta t < 20$ μs bin and, therefore, is not of the daemon origin. (By the way, the position of the -30-μs peak observed during the previous two weeks can hardly be interpreted from a similar standpoint, i.e., generation of thermal neutrons with their subsequent capture by a nucleus in the PM tube etc.)

## 5. RESULTS OF THE MARCH OBSERVATIONS

Much of what had not found ready explanation at the time was clarified in the underground experiments carried out half a year later, March 3, $16^h$, through March 26, $8.1^h$, 2006.

These experiments were conducted with the same module #3, in which the upper PM tube triggering the oscillograph viewed the 10-μm ZnS(Ag) layer on the top surface of the 4-mm-thick polystyrene plate. To protect this layer against background α radiation, it was, as before, covered by two layers of 50-μm Lavsan film. Because the film reduced strongly the α background, we believed it justified to abandon blowing the system through with liquid nitrogen vapors. This increased, however, by ~20% the fraction of NLS triggers from the upper PM tube, which is apparently due to the ZnS(Ag) scintillator responding to numerous δ electrons knocked out by the β and γ radiation of radon and its derivatives from the polystyrene underlying the ZnS(Ag) coating. Because we are interested only in HPS events from the upper PM tube, this factor, while increasing the background, can hardly affect noticeably the final conclusions (and at any rate is not capable of embellishing them). The triggering level in this experiment was lowered to ≈2.4 mV, thus increasing the trigger rate to one in ~40 s. The number of double events with NLS and $U_1 \geq 2.8$ mV on the first trace was 80 (with 14 out of them, with $|\Delta t| \leq 0.4$ μs, being caused by cosmic rays).

The results of the measurements are listed in Table 3.

Viewed in parallel with the September data discussed above, these results appear puzzling; indeed, in contrast to the September experiment, there is no maximum in the $-40 < \Delta t < -20$ μs bin. As in the ground-level March experiments [], the maximum lies in the $+20 < \Delta t < -20$ μs bin (Fig. 5), i.e., the interval corresponding to downward fall of NEACHO daemons.

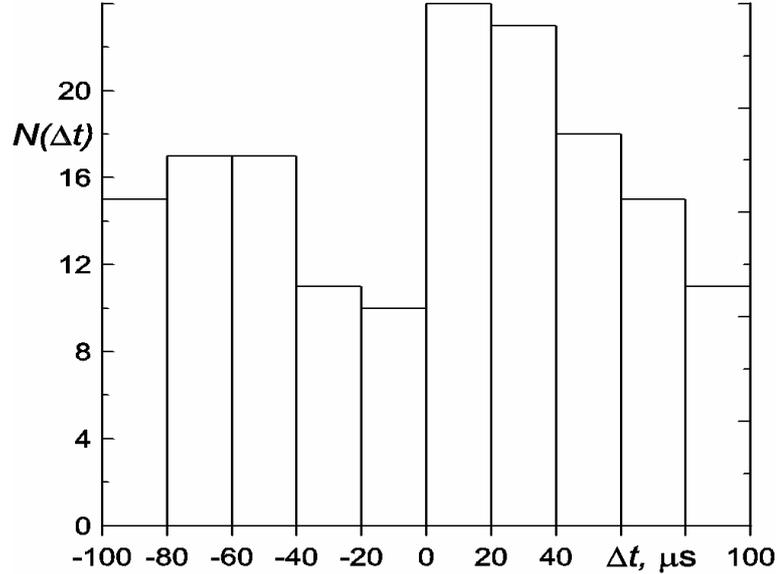

**Figure 5.** Similar as Fig.1 but for time span from March 3 ($16^h$) to March 18 ($4^h$) 2006. $N(\Delta t)$ is 15, 17, 17, 11, 10, 24, 23, 18, 15, 11; 161 events altogether (see Table 3). Statistical significance of the maximum in the +30-μs bin $\{\sigma\} = (N_{+30} - 0.1\Sigma)\cdot N_{+30}^{-1/2} = 1.44\sigma$.

Unfortunately, the maximum associated with the fall of NEACHO objects is paralleled, even to a greater relative extent than was the case in September, by a maximum in the $0 < \Delta t < 20$ μs bin. It grows with practically the same rate as the September peak. A fraction of the events making it up may certainly be caused by the fall of objects moving in the extreme outer NEACHOs, but the statistics is still too poor to allow anywhere near definite conclusions. The more so that the September experiments (see end of Sec. 4 and Fig. 4) do not bear out this assumption.

Because of competition with this near-zero maximum, the confidence level of the +30-μs peak accumulated during the whole observation period does not exceed $(1.4-1.6)\sigma$, which corresponds to only 5-7 events above background (see Fig. 5), i.e., $f_\oplus \approx 0.4 \times 10^{-7}$ cm$^{-2}$s$^{-1}$. As seen also from Fig. 6, the behavior of $N_{+30}(t)$ undergoes a break after March 18, with the rate of event arrival increasing ~1.5 times. We note also that the behavior of $N_{-30}(t)$, which corresponds to the bin recording the upward flux of the objects, reveals after March 18 likewise an increase in the rate of arrival of the events.

As a result, the $N(\Delta t)$ distribution accumulated in the period from March 18 to April 24 (see Fig. 7) has two symmetric maxima, $N_{+30}$ and $N_{-30}$; in other words, starting from March 18, the upward flux rises to catch up with the downward daemon flux.

## 6. DISCUSSION OF THE MARCH DATA

The latter observation can be interpreted as recording of two different daemon populations, more specifically, that before March 18 we record primarily objects falling from NEACHOs, which, having crossed our detector and, subsequently, the Earth, escape to reenter the NEACHOs. The slowest part of them is, however, decelerated by the Earth matter to the extent where they become capable to transfer to GESCOs and accumulate there. It is these daemons that apparently produce the second population symmetric with respect to the up/down direction, whose flux $f_⊕ ≈ 0.6×10^{-7}$ cm$^{-2}$s$^{-1}$ is initially recorded in the +30 and -30 µs bins, but after accumulation of the objects in GESCOs increases to slightly exceed the flux of the primary "through" population ($f_⊕ ≈ 0.4×10^{-7}$ cm$^{-2}$s$^{-1}$).

As already mentioned, the appearance of the maximum in the -30-µs bin of the September $N(\Delta t)$ distribution came to us as a surprise. We first supposed that it resulted from the scintillator plates having been turned with the ZnS(Ag) layer up (this was caused by the need to place two Lavsan films on ZnS(Ag), see above). Indeed, we used earlier (Drobyshevski *et al.*, 2003a) the dependence of the HPS shape on the direction in which the daemon that has captured an atomic nucleus in the ZnS(Ag) layer was moving, into the air or into the bulk of the polystyrene. In the latter case, part of the nucleons emitted in the capture by a nucleus being dragged by a daemon stop in the polystyrene and do not reach the phosphor, thus distorting the scintillation pulse shape. In our case, we made an attempt to assign the appearance of the -30 µs maximum to the quasi-vacuum properties of polystyrene. We assumed that an active daemon (i.e., the complex of the remainder of the nucleus captured earlier plus the daemon carrying a negative charge $|Z_n| - |Z| < 0$) does not capture a carbon nucleus or protons because of the difficulties it would be faced in getting rid of excess energy (indeed, the excitation energy of the first nuclear level of carbon is 4.4 MeV) and, therefore, only on entering the ZnS(Ag) layer would it be able to capture a new nucleus by expending the excess capture energy in its excitation (for the Zn nucleus, the excitation energy is ~1 MeV).

**Table 3**. $N(\Delta t)$ distributions (-100 < $\Delta t$ < 100 µs) accumulated from March 3 (16$^h$00') to March 26 (6$^h$42') 2006. Ten bins, each 20 µs wide, are centered at the $\Delta t$ specified in the Table. The $N(\Delta t)$ distributions are given for the time of recording of an event in the +30-µs bin. $\Sigma$ is the total number of events in $N(\Delta t)$; $N_{+30}$ is the number of events in the +30-µs bin; $\{\sigma\} = (N_{+30} - 0.1\Sigma) \cdot N_{+30}^{-1/2}$ refers to statistical significance of the +30-µs maximum; $\{N\}_9 = (\Sigma - N_{+30})/9$ is the average number of 'background' events per one bin (excepting the +30 µs bin); $\{N\}_8 = (\Sigma - N_{+10} - N_{+30})/8$ is the average number of 'background' events per one bin (excepting the +10 and +30 µs bins).

| Event № | March days | -90 | -70 | -50 | -30 | -10 | 10 | 30 | 50 | 70 | 90 | $\Sigma$ | $\{\sigma\}$ | $\{N\}_9$ | $\{N\}_8$ |
|---|---|---|---|---|---|---|---|---|---|---|---|---|---|---|---|
| 1161500 | 3,667 | | | | | | | | | | | | | | |
| 1162569 | 4,402 | 2 | 0 | 0 | 0 | 3 | 0 | 1 | 1 | 2 | 0 | 9 | 0,10 | 0,89 | 1,00 |
| 1163748 | 4,748 | 2 | 0 | 0 | 0 | 3 | 0 | 2 | 1 | 2 | 0 | 10 | 0,71 | 0,89 | 1,00 |
| 1163997 | 4,944 | 3 | 0 | 0 | 0 | 3 | 1 | 3 | 1 | 2 | 0 | 13 | 0,98 | 1,11 | 1,12 |
| 1165561 | 6,065 | 3 | 4 | 3 | 0 | 3 | 2 | 4 | 4 | 4 | 2 | 29 | 0,55 | 2,78 | 2,88 |
| 1165729 | 6,190 | 3 | 5 | 3 | 0 | 3 | 3 | 5 | 4 | 4 | 3 | 33 | 0,76 | 3,11 | 3,12 |
| 1165829 | 6,271 | 3 | 5 | 3 | 0 | 3 | 3 | 6 | 5 | 4 | 3 | 35 | 1,02 | 3,22 | 3,25 |
| 1166113 | 6,494 | 3 | 5 | 4 | 0 | 3 | 4 | 7 | 6 | 4 | 3 | 39 | 1,17 | 3,56 | 3,50 |
| 1166981 | 7,141 | 3 | 5 | 5 | 2 | 4 | 4 | 8 | 7 | 4 | 4 | 46 | 1,20 | 4,22 | 4,25 |
| 1168207 | 8,026 | 4 | 5 | 6 | 3 | 6 | 5 | 9 | 8 | 4 | 7 | 57 | 1,10 | 5,33 | 5,38 |
| 1170307 | 8,916 | 4 | 6 | 9 | 4 | 7 | 7 | 10 | 10 | 5 | 7 | 69 | 0,98 | 6,56 | 6,50 |
| 1170352 | 8,941 | 4 | 6 | 9 | 4 | 7 | 7 | 11 | 10 | 5 | 7 | 70 | 1,21 | 6,56 | 6,50 |
| 1170489 | 9,026 | 4 | 6 | 9 | 4 | 7 | 7 | 12 | 10 | 5 | 7 | 71 | 1,41 | 6,56 | 6,50 |
| 1172827 | 10,621 | 4 | 10 | 10 | 5 | 8 | 10 | 13 | 12 | 6 | 8 | 86 | 1,22 | 8,11 | 7,88 |
| 1176863 | 12,635 | 7 | 12 | 11 | 6 | 9 | 14 | 14 | 13 | 9 | 9 | 104 | 0,96 | 10,00 | 9,50 |
| 1176976 | 12,699 | 7 | 12 | 11 | 6 | 9 | 14 | 15 | 14 | 9 | 9 | 106 | 1,14 | 10,11 | 9,62 |
| 1177778 | 13,216 | 9 | 12 | 12 | 7 | 10 | 16 | 16 | 14 | 9 | 9 | 114 | 1,15 | 10,89 | 10,25 |
| 1179308 | 14,297 | 10 | 14 | 13 | 8 | 10 | 19 | 17 | 16 | 9 | 10 | 126 | 1,07 | 12,11 | 11,25 |
| 1179352 | 14,318 | 10 | 14 | 13 | 8 | 10 | 19 | 18 | 16 | 9 | 10 | 127 | 1,25 | 12,11 | 11,25 |
| 1181214 | 15,457 | 11 | 14 | 14 | 8 | 10 | 19 | 19 | 16 | 13 | 11 | 135 | 1,26 | 12,89 | 12,12 |
| 1182018 | 15,960 | 12 | 14 | 15 | 8 | 10 | 21 | 20 | 16 | 13 | 11 | 140 | 1,34 | 13,33 | 12,38 |
| 1183482 | 16,884 | 13 | 15 | 16 | 9 | 10 | 21 | 21 | 16 | 14 | 11 | 146 | 1,40 | 13,89 | 13,00 |
| 1185456 | 17,883 | 15 | 17 | 17 | 11 | 10 | 24 | 22 | 18 | 15 | 11 | 160 | 1,28 | 15,33 | 14,25 |
| 1185549 | 17,913 | 15 | 17 | 17 | 11 | 10 | 24 | 23 | 18 | 15 | 11 | 161 | 1,44 | 15,33 | 14,25 |
| 1189016 | 19,391 | 15 | 21 | 20 | 14 | 11 | 27 | 24 | 20 | 19 | 11 | 182 | 1,18 | 17,56 | 16,38 |
| 1189068 | 19,412 | 15 | 21 | 20 | 15 | 11 | 27 | 25 | 20 | 19 | 11 | 184 | 1,32 | 17,67 | 16,50 |
| 1189147 | 19,448 | 15 | 21 | 20 | 15 | 11 | 27 | 26 | 20 | 19 | 11 | 185 | 1,47 | 17,57 | 16,50 |
| 1189510 | 19,617 | 15 | 21 | 21 | 16 | 11 | 27 | 27 | 20 | 19 | 11 | 188 | 1,58 | 17,81 | 16,75 |
| 1189942 | 19,796 | 15 | 21 | 21 | 17 | 13 | 28 | 28 | 20 | 19 | 11 | 193 | 1,64 | 18,33 | 17,12 |
| 1191974 | 20,600 | 17 | 22 | 24 | 19 | 14 | 31 | 29 | 22 | 20 | 12 | 210 | 1,48 | 20,11 | 18,75 |
| 1192918 | 20,998 | 17 | 22 | 24 | 19 | 15 | 31 | 30 | 22 | 20 | 13 | 213 | 1,59 | 20,33 | 19,00 |
| 1199835 | 23,234 | 18 | 26 | 28 | 22 | 17 | 33 | 31 | 25 | 23 | 17 | 240 | 1,26 | 23,22 | 22,00 |
| 1200213 | 23,410 | 18 | 27 | 28 | 22 | 17 | 34 | 32 | 25 | 23 | 18 | 244 | 1,34 | 23,56 | 22,25 |
| 1201168 | 23,868 | 18 | 27 | 28 | 23 | 17 | 34 | 33 | 25 | 24 | 19 | 248 | 1,43 | 23,89 | 22,62 |
| 1201630 | 24,063 | 18 | 27 | 28 | 23 | 17 | 34 | 34 | 25 | 24 | 19 | 249 | 1,56 | 23,89 | 22,62 |
| 1202457 | 24,406 | 18 | 28 | 28 | 24 | 17 | 34 | 35 | 25 | 24 | 19 | 252 | 1,66 | 24,11 | 22,88 |
| 1202603 | 24,467 | 19 | 28 | 28 | 24 | 17 | 34 | 36 | 25 | 24 | 19 | 254 | 1,77 | 24,22 | 23,00 |
| 1206942 | 26,279 | 21 | 31 | 31 | 26 | 19 | 36 | 36 | 26 | 25 | 19 | 270 | 1,50 | 26,00 | 24,75 |

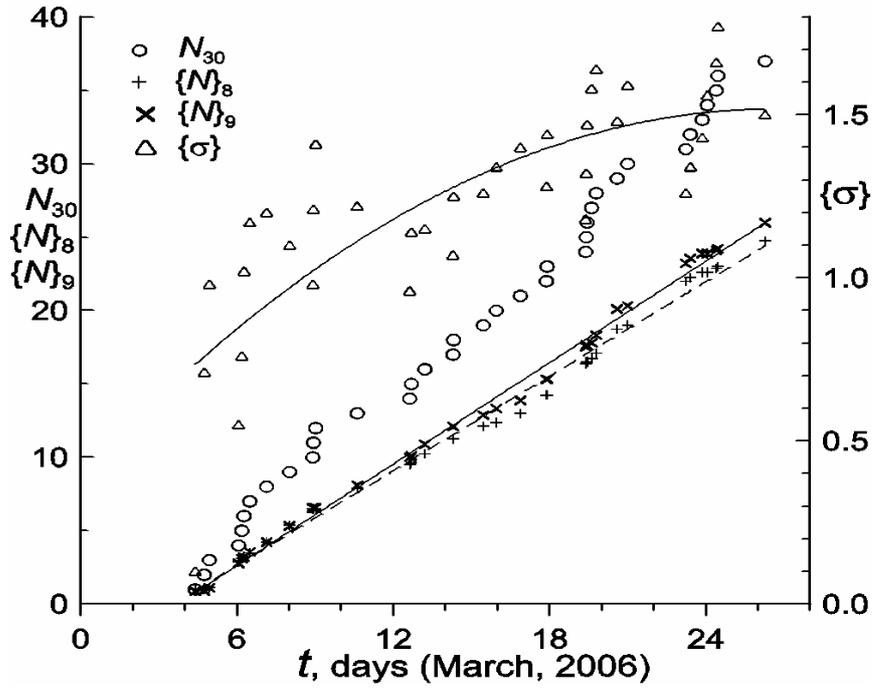

**Figure 6.** The same as Fig. 2 but for March 2006. $N_{+30}(t)$ is the number of events in the +30-µs bin of the $N(\Delta t)$ distribution. $\{N(t)\}_9 = 1.15 \cdot t - 4.29$ approximation has a C.L. = 99.7%; $\{N(t)\}_8 = 1.07 \cdot t - 3.81$ (C.L. = 99.6%); $\{\sigma\}$ is statistical significance of the +30-µs bin maximum in the $N(\Delta t)$ distribution.

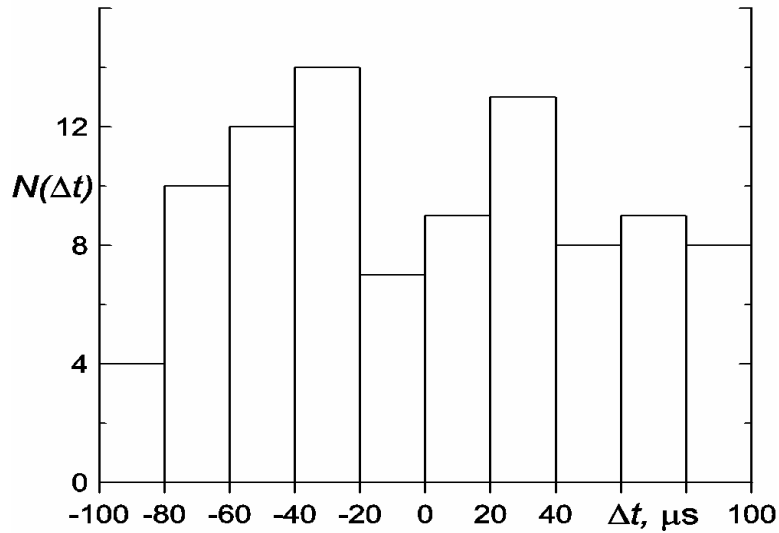

**Figure 7.** The same as Figs. 1, 4, and 5 but for the time span from 18 ($4^h$) to 24 ($24^h$) March 2006 ($N(\Delta t)$ is: 4, 10, 12, 14, 7, 9, 13, 8, 9, 8 ; 94 events altogether).

Therefore, when starting our observations at BNO in March 2006, we were expecting to find the maximum in the -30 μs bin. The maximum appeared, however, in the +30 μs bin, which was quite unexpected, although in ground-level experiments with ZnS(Ag) deposited on the lower side of the polystyrene plate the March maximum always appeared in the +30 μs bin.

One has thus to find an explanation to the observation that in March daemons fall from above and produce a peak in the +30 μs bin, whereas in September they arrive in an upward-moving stream to be recorded in the -30 μs bin.

The situation becomes clearer if we recall the celestial mechanics aspects of the problem. The Earth's axis of rotation is inclined by 23.5° to the ecliptic plane, which accounts, as is well known, for the alternation of seasons. Our detector responds to objects with a vertical velocity component. It is located in the Northern hemisphere. In March, the Northern hemisphere of the Earth is inclined backward relative to the direction of its orbital motion. In September, it is oriented forward. Therefore, the observed effect should appear if daemons catch up with the Earth, in particular, in moving in NEACHOs lying predominantly outside the Earth's orbits, in particular, in those with perihelia close to the Earth's orbit.

The objects catching up with the Earth with velocities of up to ~12.3 km s$^{-1}$ appear in these orbits as a result of their gravitational interaction with the Earth. It is known that a small body at rest colliding elastically with a moving body of a large mass can acquire a velocity twice that of the large body. Coulomb (gravitational) interaction is elastic, so that gravitational interaction of daemons with the Earth should initially first transfer them and accumulate in outer NEACHOs with respect to the Earth. The escape velocity for the Earth being 11.19 km s$^{-1}$, it cannot change the velocity of oncoming particles by more than this value. For a body to escape from the Earth's orbit to infinity, its velocity should exceed the orbital velocity of the Earth of 29.8 km s$^{-1}$ by 12.3 km s$^{-1}$. It thus becomes clear that particles can escape from NEACHOs to infinity only through gradual summation of the gravitational perturbations imparted to them by the Earth and other planets. This process by which daemons increase their energy in NEACHOs is counteracted by the slowing down by matter in their rare transits through the Earth, as a result of which their orbits approach that of the Earth. The net outcome of these processes would require a comprehensive analysis. Nevertheless, there are grounds to believe that daemons should accumulate in outer orbits with respect to the Earth, which have a noticeable eccentricity and approach closely at perihelion the Earth's orbit on the outside. Such orbits could be called NEACHOs only with some reservations. Objects leaving them would fall primarily on the night side of the Earth, or rather late in the evening, because they catch up with the Earth. This relates also to the March flux of the through population moving downward. Because in September we record daemons that have crossed the Earth, their upward flow should reach a maximum in the morning. And only the few daemons that were slowed down by the Earth's matter, transferred into GESCOs and accumulated in them will enter the detector both from above and from below (see Fig. 7). Although the relevant statistics is poor, our observations (Fig. 8) are not at odds with expectations of the presence of such diurnal variations in the flux of the "through" population.

There should exist also a latitude dependence of these daemon fluxes, which may manifest itself, if in nothing else, in the absence of a clearly pronounced -30 μs maximum in the September measurements conducted at St. Petersburg (60° latitude). By contrast, in the Southern hemisphere we should have an opposite situation. The March maximum, in particular, should be seen there in the -30 μs bin and originate from upward-moving, "through" objects in the first half of the day. This stresses the importance of performing simultaneous measurements in different latitudes (the latitude of BNO is 43.2°) and, of course, in the Southern hemisphere.

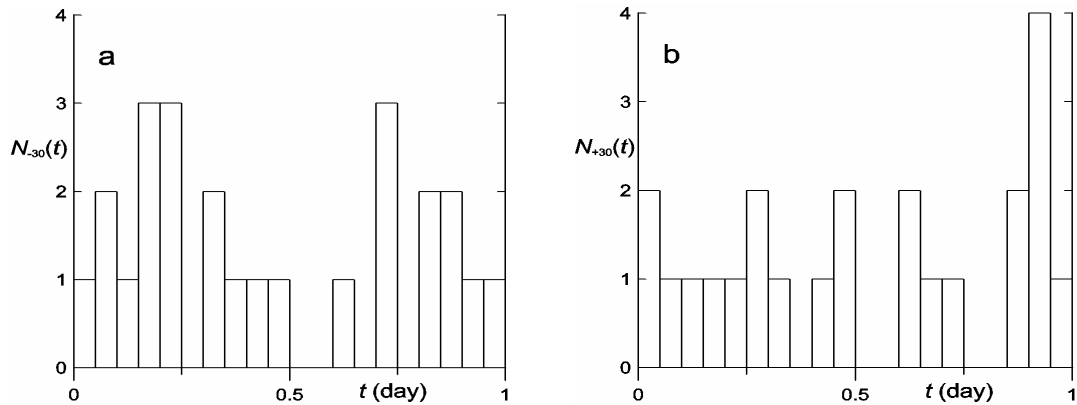

**Figure 8**. Distribution of double events vs. time of day (reckoned from midnight, day is divided into 20 intervals): (a) events in -30 μs bin, September 3 through 16, 2005 (25 events altogether: 1, 2, 1, 3, 3, 0, 2, 1, 1, 1, 0, 0, 1, 0, 3, 0, 2, 2, 1, 1); (b) events in +30 μs bin, March 3 through 18, 2006 (23 events altogether: 2, 1, 1, 1, 1, 2, 1, 0, 1, 2, 0, 0, 2, 1, 1, 0, 0, 2, 4, 1).

It is appropriate to note here one more point. We have in mind the specific features of gravitational focusing of the daemon flux by the Earth. The total flux of daemons catching up with the Earth and striking it in March in the Northern hemisphere from above is amplified by the Earth's gravitational pull nearly twofold, but it maintains monotony of its latitude (and daytime) distribution, because outside the Earth (until it enters the detector), the flow feels gravitational attraction, as it were, of a point mass. The September flow of "primary" daemons which have crossed the Earth was subjected there to gravitation of a quasi-spherical, but radially non-uniform mass. Focusing of this kind may give rise to formation of surfaces (caustics) with an enhanced (or lower) flux. Therefore, one may envisage considerable temporal and qualitative differences in variation of the parameters of the March and, particularly, September primary "through" fluxes. Because of the rosette-like motion in the field of a non-point mass, the secondary population trapped into GESCOs forgets its original orientation, a factor that should bring about disappearance of the diurnal (and latitude) dependences of its flow through the Earth's surface.

## 7   MAIN CONCLUSIONS

The low-background experiments intentionally performed at the BNO in September 2005 and in March 2006 and aimed at observation of the expected maxima in the low-velocity daemon flux have confirmed the existence of the maxima at the predicted time, thus providing support for the daemon paradigm. These experiments revealed also the influence of a number of previously uncontrolled parameters on the efficiency of the scintillation-based daemon detection method.

We have, for instance, confirmed the possibility of using some FEU-167-1 PM tubes as sensitive elements; indeed, daemons crossing the PM evacuated envelope and carrying the remainder of a previously captured nucleus retain and even increase their net negative charge, so that such a PMT detects the passage of daemons with an efficiency of tens of percent.

The working hypothesis by which for the base distance needed for the velocity estimates in St.Petersburg experiments with double scintillators was chosen not the 7-cm gap between them but rather the 29-cm separation between the top scintillator plate and the lower tinned-iron sheet of the module case (Drobyshevski 2002) has found support. Initially, this hypothesis was based on celestial mechanics considerations that the daemons striking the Earth should accumulate in NEACHOs and have a velocity of not less than 11.2 km s$^{-1}$, which accounts for the small dispersion of their velocities (~11.2-15 km s$^{-1}$) making the 30-μs peak in the March $N(\Delta t)$ distribution so narrow and revealing.

It has been shown that the uncontrollable (and varying) radon background (as well as differences in not photometric but rather design parameters of PM tubes) could be a major cause of poor reproducibility of the ground-level experiments performed at the PTI in 2002-2005. Indeed, in March 2000 the observations were run in a well ventilated room 85 m$^3$ in volume, and starting with June 2002, in a 43 m$^3$ closed room with a split-type air conditioner. The measurements of the HPS/α background carried out already in December 2005 with module #1 (see Paper I) reassembled in each room and run for several hours, thus excluding the decrease in radon concentration due to its decay inside the module, showed the HPS background in the second room to be 4-5 times higher than that in the first one (this difference depends on the direction and strength of the wind, which appears only natural in the conditions where radon enters the air from the walls or the soil; see, e.g., Pertsov (1964)).

By using the FEU-167-1 PM tubes as a high-efficiency vacuum sensor responding to daemon passage through them, we have succeeded in raising hundredfold the detector efficiency per unit area. Indeed, in March 2000 the 1-m$^2$ detector recorded 18 excess events in the 30-μs bin in 2.5×10$^6$ s, which corresponds to a flux ≈ 0.7×10$^{-9}$ cm$^{-2}$s$^{-1}$. In the Baksan experiment performed in September 2005, the dia. 12 cm PM tube (115 cm$^2$ area) recorded 8-9 excess events in 0.7×10$^6$ s, thus yielding $f_\oplus$ ≈ 1.0×10$^{-7}$ cm$^{-2}$s$^{-1}$. Allowing for the low efficiency of the scintillator section of the detector (~3%), the true value of the maximum daemon flux from the NEACHOs may be 10-30 times higher, a figure which fits fairly well our old estimates (Drobyshevski 1997).

Because the dimensions of the lower sensor in these experiments were reduced compared with the original ground-level experiments from 50 to 10 cm, the solid angle viewed by the detector decreased approximately to one half. It is the increase in angular resolution that apparently has resulted in our having revealed finer details in the daemon flux variation, in particular, some evidence for the existence of diurnal variations, as well as of a primary (in NEACHOs) and a secondary (in GESCOs) populations. It becomes understandable now why in the measurements of the year 2000 the March maximum extended from February 27 to March 27 [], whereas this year it is shorter and lasts approximately from March 3 to 24.

One has naturally to carry out a comprehensive theoretical analysis of the capture of daemons from the Galactic disc into NEACHOs and their evolution in these orbits, with subsequent transfer of a part of them into GESCOs, etc. Investigation of the latitude and diurnal variations would require accumulation of reliable statistics, including parallel experiments in the Northern and Southern hemispheres.

To sum up, our Baksan experiments aimed at detection of the September and March daemon fluxes have revealed a number of factors that had not been controlled properly in previous ground-level and high latitude experiments in St.Petersburg. The underground experiments permitted measurements with detector of a new type developed with a deeper understanding of daemon behavior in matter and in vacuum.

The results of these observations provide compelling evidence for the many consequences and conclusions of the daemon paradigm. We believe that the most fundamental and intriguing implication of the DM in the Universe being made of Planckian rather than other (e.g., of the type of WIMPs) objects is possibly that each of them, in principle, can be an independent quasi-closed universe very weakly opened by a small electric charge (coined "friedmon" by Markov and Frolov, 1970). We approach here realization of a concept of Multi-Universe Cosmos (see, e.g., Velan, 1992), where each universe contains countless similar universes (and so, possibly, *ad infinitum*).

**ACKNOWLEGEMENT**

The authors are deeply indebted to Dr V.V. Kuzminov, Director of BNO, for invaluable advises and help in organizing the experiment in BNO. Sincere thanks are due also to Prof A.G. Zabrodski, Director of Ioffe Institute, for his appreciation of the fundamental significance of the work and for its invariable support.

**REFERENCES**

Drobyshevski E.M. (1997) If the Dark Matter objects are electrically multiply charged: New opportunities. In: Klapdor-Kleingrothaus and H.V., Ramachers Y., eds., *Dark Matter in Astro- and Particle Physics*, World Scientific, Singapore, pp.417-424.

Drobyshevski E.M. (2000) Time of explosive decay of daemon-containing nucleus. *Mon. Not. Roy. Astron. Soc.*, **311**, L1-L3.

Drobyshevski E.M. (2002) Detecting the dark electric matter objects (daemons). *A&ATrans.,* **21**, 65-73.

Drobyshevski E.M. (2005) Detection of Dark Electric Matter Objects falling out from Earth-crossing orbits. In: Spooner N.J.C., Kudryavtsev V., eds., *Proc. of the 5th Intnl. Workshop on "The Identification of Dark Matter"*, World Scientific, Singapore, pp.408-413.

Drobyshevski E.M. (2006) Study of the spring and autumn daemon flux maxima at the Baksan Neutrino Observatory. Preprint *astro-ph*/0605314; *A&ATrans.* (in press) (Paper I).

Drobyshevski E.M., Beloborodyy M.V., Kurakin R.O., Latypov V.G., and Pelepelin K.A. (2003a) Detection of several daemon populations in Earth-crossing orbits. *A&ATrans.*, **22**, 19-32.

Drobyshevski E.M., Drobyshevski M.E., Izmodenova T.Yu. and Telnov D.S. (2003b) Detection of several daemon populations in Earth-crossing orbits. *A&ATrans.*, **22**, 263-271.

Markov M.A. (1966) Elementary particles with largest possible masses (quarks and maximons). *ZhETF,* **51**, 878-890.

Markov M.A. and Frolov V.P. (1970) Metrics of the closed Friedman world perturbed by electric charge (to the theory of electromagnetic "friedmons"). *Teor. Mat. Fiz.*, **3**(1), 3-17 (in Russian).

Pertsov L.A. (1964) *The Natural Radioactivity of Biosphere*, Atomizdat, Moscow (in Russian).

Taylor J. R. (1982) *An Introduction to Error Analysis*, Univ. Science Books, Mill Valley, CA.

Velan A.K. (1992) *The Multi-Universe Cosmos*, Plenum Press, N.-Y. and London.